\documentclass[twocolumn,showpacs,preprintnumbers,amsmath,amssymb]{revtex4}

\usepackage{graphicx}
\usepackage{dcolumn}
\usepackage{bm}

\begin{document}

\title{Reconciling long-term cultural diversity and short-term collective social behavior} 

\author{Luca Valori$^{1}$, Francesco Picciolo$^{1}$, Agnes Allansdottir$^{2}$, Diego Garlaschelli$^{3}$}

\affiliation{
$^1$Department of Chemistry, University of Siena, Via A. De Gasperi 2, 53100 Siena, Italy\\
$^2$University of Siena, 53100 Siena, Italy\\
$^3$Lorentz Institute for Theoretical Physics, University of Leiden, Niels Bohrweg 2, NL-2333 CA Leiden, The Netherlands}
\date{\today}

\begin{abstract} 
An outstanding open problem is whether collective social phenomena occurring over short timescales can systematically reduce
cultural heterogeneity in the long run, and whether offline and online human interactions contribute differently to the process.
Theoretical models suggest that short-term collective behavior and long-term cultural diversity are mutually excluding, since they require very different levels of social influence.  
The latter jointly depends on two factors: the topology of the underlying social network and the overlap between individuals in multidimensional cultural space. 
However, while the empirical properties of social networks are well understood, little is known about the large-scale organization of real societies in cultural space, so that random input specifications are necessarily used in models.
Here we use a large dataset to perform a high-dimensional analysis of the scientific beliefs of thousands of Europeans. 
We find that inter-opinion correlations determine a nontrivial ultrametric hierarchy of individuals in cultural space, a result unaccessible to one-dimensional analyses and in striking 
contrast with random assumptions.
When empirical data are used as inputs in models, we find that ultrametricity has strong and counterintuitive effects, especially in the extreme case of long-range online-like interactions bypassing social ties. 
On short time-scales, it strongly facilitates a symmetry-breaking phase transition triggering coordinated social behavior. 
On long time-scales, it severely suppresses cultural convergence by restricting it within disjoint groups.
We therefore find that, remarkably, the empirical distribution of individuals in cultural space appears to optimize the coexistence of short-term collective behavior and long-term cultural diversity, which can be realized simultaneously for the same moderate level of mutual influence.
This also shows that long-range interactions may simultaneously enhance coordination and sustain diversity.
\end{abstract}

\pacs{Valid PACS appear here}

\maketitle

\section{Introduction}
How a society spontaneously organizes macroscopically from the microscopic, uncoordinated behavior of individuals is one of the most studied and exciting problems of modern science \cite{socialatom,castellanoreview,computationalsocialscience}. 
Collective social phenomena are observed in different aspects of everyday life, including the onset of large-scale popularity and fashion (both offline \cite{fortunato2007scaling,Fowlercascades} and online \cite{onlinepopularity,felix_online}), the existence of large fluctuations and herding behavior in financial markets \cite{mantegna,contbouchaud,econophysics}, the spontaneous emergence of order in traffic and crowd dynamics \cite{sociophysics}, 
the properties of voting dynamics \cite{fortunato2007scaling,galam},
the structure of country-wide communication networks \cite{phonecalls,footprints,computationalsocialscience}, the spreading of habits, fear, gossip, rumors, etc. \cite{castellanoreview,sociophysics}.
When collective phenomena occur, large parts of a population turn out to be globally correlated as a result of the combination of many local interactions, even if no centralized mechanism takes place. 
Importantly, the collective outcome is different from a mere superposition of non-interacting individual behaviors, and contrasts the `representative agent' scenario often postulated in economic theories.
The different characteristics, choices and behaviors of individuals, rather than being `averaged out' in the long run and at a large scale, may in some circumstances become amplified at the societal level \cite{socialatom}. 
The observation of these phenomena, also enabled recently by large-scale electronic platforms where people can exchange information with unprecedented speed and breadth (see Appendix for a more detailed discussion), poses the question of whether the diversity of behaviors, attitudes and opinions is destined to be progressively reduced in the long run. Naively, one expects that stronger collective social phenomena taking place on short timescales may gradually result in more homogeneous behaviors in the long term. 

This picture is reinforced by the fact that similar mechanism are believed to be among the key driving forces of both collective social behavior \cite{socialatom,felix_online} and cultural convergence \cite{axelrod}.
Various simplified models have been introduced to quantitatively simulate the fate of cultural diversity and the dynamics of opinions in large groups \cite{castellanoreview,sociophysics,galam}. In both cases, the main hypothesized mechanisms are the tendency of social interactions to favor convergence and consensus (\emph{social influence} \cite{Festinger}) and the inverse tendency of culturally similar individuals to interact more than dissimilar ones (\emph{homophily}) \cite{axelrod}. 
Recently, the concept of homophily has been enriched with the quantitative notion of \emph{bounded confidence}, according to which people are not culturally influenced by too dissimilar peers \cite{Flache_and_Macy_2006,De_Sanctis_and_Galla_2007,castellanoreview}. 
Quantitatively, when the opinions or cultural traits of an individual are represented as a scalar or vector variable, the bounded confidence hypothesis results in two individuals being potentially influenced by each other only if the distance between their associated variables is smaller than a certain threshold, representing the level of confidence or \emph{tolerance} \cite{Flache_and_Macy_2006,De_Sanctis_and_Galla_2007,castellanoreview}. 
This threshold, which in a simplified picture is assumed to have the same value $\omega$ across the entire population, quantifies how susceptible an individual is to possible cultural influences. 
According to this picture, two individuals can only influence each other if they are socially tied and also sufficiently similar culturally: the effective medium of interaction is the overlap between the social network and the \emph{cultural graph} connecting pairs of similar individuals (see Appendix for an extended discussion).
Rather than conveying a detailed picture of reality, models of social dynamics aim at understanding the effects that different mechanisms proposed in social science may have when combined together and when taking place at a large-scale level. In particular, the importance of one of the most popular models, proposed by Axelrod \cite{axelrod}, resides in showing that social influence and homophily do not necessarily reinforce each other and determine a culturally homogeneous society. In fact, the model suggests that the persistence of cultural diversity in the long term is warranted by the inhibition of influence among dissimilar individuals, even if socially tied. However, if plugged into other models of social dynamics \cite{contbouchaud,castellanoreview,sociophysics}, the same mechanism prevents information diffusion across culturally disconnected groups, and therefore also implies no collective social behavior in the short term. 
We will give an explicit example of this effect simulating both short- and long-term dynamics on random data.

According to the above results, the coexistence of long-term cultural diversity and short-term collective behavior is apparently a paradox, which can only be solved by invoking different mechanisms at different timescales. 
However, here we show that, even without postulating more complicated scenarios, the paradox can be explained by taking into account an insofar ignored aspect of empirical  multidimensional cultural profiles. 

\section{The hierarchical distribution of individuals in cultural space}
Our study starts with the analysis of the large \emph{Eurobarometer} dataset\cite{agnes1,agnes2}, an official report of a questionnaire-based survey\cite{dataset,eurobarometer} of the European Commission, which allowed us to reconstruct the empirical multidimensional vectors of 13,000 individuals across 12 European countries. 
In the Appendix we describe the dataset in more detail, and how the multiple-choice nature of the questionnaire allowed us to define, for each individual $i$ in country $\alpha$, an $F$-dimesional vector $\vec{v}_i$ whose $k$th component $v_i^{(k)}$ represents the answer given by $i$ to question $k$ in the survey ($k=1,\dots F$, where the number of questions is $F=161$). To obtain groups of equal size, we sampled $N=500$ individuals for each of the 12 countries, plus a thirteenth group of $N=500$ individuals sampled all across Europe. We labeled each group with a Greek letter $\alpha=1,\dots 13$.
For each pair $i,j$ of individuals we defined a normalized metric distance $d_{ij}^{(k)}$ ($0\le d_{ij}^{(k)}\le 1$) between $v_i^{(k)}$ and $v_j^{(k)}$, measuring how different the answers given by $i$ and $j$ to question $k$ were. We also defined an overall metric distance $d_{ij}$ (again such that $0\le d_{ij}\le 1$) between the entire sets of answers $\vec{v}_i$ and $\vec{v}_j$ given by $i$ and $j$ (see Appendix for all 
definitions).

In addition to real data, we considered two types of randomized data which represent important null models providing informative benchmarks throughout our analysis.
A first type of randomization (`random answers') simply consists in defining $N$ random vectors, each obtained drawing $F$ answers uniformly among the possible alternatives. This simulates $N$ individuals giving completely random answers to the questionnaire, and does not depend on the empirically observed answers. This provides a  unique random benchmark against which all sampled groups can be compared, and corresponds to the usual initial specification in the Axelrod model \cite{axelrod} and similar models \cite{Flache_and_Macy_2006,De_Sanctis_and_Galla_2007,castellanoreview}.
A second type of randomization (`shuffled answers') consists in randomly shuffling, for each of the $F$ questions in the questionnaire, the real answers given by the $N$ individuals of the group considered. In this case, different groups have different randomized benchmarks, each characterized by its own probability distribution (determined by real data) of possible answers. This null model is very important, as it preserves the number of times a particular answer was actually given to each question (so it preserves `more fashioned' answers for each group), but destroys the correlations between answers given by the same individual to different questions.
\footnote{Note that in more traditional one-dimensional analyses where only a single opinion is considered, the shuffling procedure is impossible: shuffled data are equivalent to the original data, and the only possible null model would be the first one we described (or one where the empirical abundances are modified arbitrarily).}

We analyzed several properties of real and randomized data, as a preliminary step before studying the impact of the empirical structure on short- and long-term dynamics. 
A peculiar aspect of our multidimensional data is the possibility to investigate 
cross-correlations among opinions, an information which is not available in 
one-dimensional studies. 
In particular, we studied whether small (large) differences between the 
answers given to question $k$ imply small (large) differences between answers 
to question $l$ by measuring the correlation $\rho^{(kl)}_\alpha$ between $d_{ij}^
{(k)}$ and $d_{ij}^{(l)}$ for all pairs $k,l$ of questions and for all pairs $i,j\in \alpha$ of individuals belonging to group $\alpha$ (see Appendix).
We found that, whereas random and shuffled data display no significant correlation by construction, real data are always characterized by a predominance of strong positive correlations, plus a minority of weak negative correlations.
This pattern is analogous to the `likes attract' phenomenon: individuals with more beliefs in common are more likely to agree on other opinions (strong positive correlation), while dissimilar individuals tend to ignore, rather than repel, each other (weak negative correlation).
However, in this case we have an evidence of a deeper mechanism, since we know that individuals in our data are socially unrelated. Therefore, rather than an effect of homophily and social influence, the observed result is the signature of intrinsic inter-opinion correlations in a single individual.

\begin{figure}
\begin{center}
\includegraphics[width=.45\textwidth]{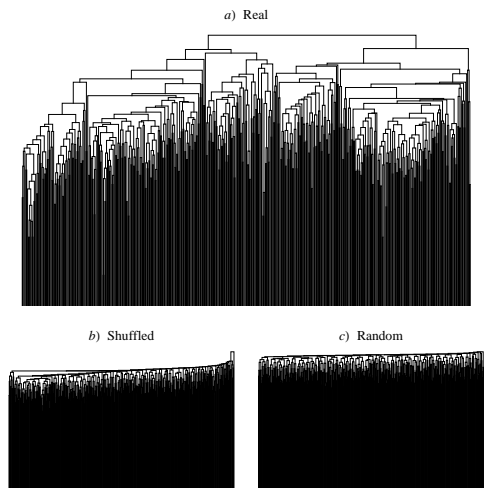}
\end{center}
\caption{Dendrograms resulting from the application of an average linkage clustering algorithm to the cultural vectors $\{\vec{v}_i\}$, represented as leaves of the tree along the horizontal axis. \textbf{a)} Real Germany data. \textbf{b)} Shuffled Germany data. \textbf{c)} Random data.
\label{fig_dendro}}
\end{figure}

The observed inter-opinion correlations have important effects on the distribution of individuals, i.e. of the vectors $\{\vec{v}_i\}$, in cultural space. 
We find  (see Table \ref{tab}) that the average inter-individual distance $\mu_\alpha\equiv\langle d_{ij}\rangle_{ij\in\alpha}$ of random data is larger than real data, while it is easy to show theoretically (and confirm by looking at the measured values) that real and shuffled data always have the same value of $\mu_\alpha$, i.e. $\mu_{\alpha,real}=\mu_{\alpha,shuffled}$.
This means that the observed `attraction' among opinions does not imply, as one would naively expect, that the empirical vectors $\{\vec{v}_i\}$ are closer to each other in cultural space than shuffled data.
However, real and shuffled data differ significantly in other properties of the distribution of vectors in cultural space.
A first difference is that real distances are much more broadly distributed than shuffled ones. 
This can be inspected by measuring the intra-group variance $\sigma^2_\alpha\equiv
\langle d_{ij}^2\rangle_{ij\in\alpha}-\langle d_{ij}\rangle^2_{ij\in\alpha}$. 
As can be seen from Table \ref{tab}, $\sigma_{\alpha,real}$ is roughly twice as large as, and more variable than, $\sigma_{\alpha,shuffled}$ (see Appendix for more details).
\begin{table}
\begin{center}
\begin{tabular}{llcccc}
\hline
$\alpha$&Country&$\mu_{\alpha,real}$&$\mu_{\alpha,shuffled}$&$\sigma_{\alpha,real}$&$\sigma_{\alpha,shuffled}$\\
\hline
1&Belgium&0.299&0.299&0.052&0.024\\
2&Denmark&0.307&0.307&0.054&0.025\\
3&France&0.303&0.303&0.050&0.025\\
4&Germany&0.295&0.295&0.054&0.024\\
5&Greece&0.286&0.286&0.059&0.024\\
6&Ireland&0.295&0.295&0.057&0.024\\
7&Italy&0.295&0.295&0.052&0.024\\
8&Luxembourg&0.304&0.304&0.051&0.024\\
9&Netherlands&0.297&0.297&0.051&0.025\\
10&Portugal&0.287&0.287&0.067&0.024\\
11&Spain&0.298&0.298&0.058&0.024\\
12&UK&0.292&0.292&0.053&0.024\\
13&Europe&0.304&0.304&0.053&0.025\\
\hline
14&Random&0.445&0.445&0.027&0.027\\
\hline
\end{tabular}
\end{center}
\caption{Average ($\mu_\alpha$) and standard deviation ($\sigma_\alpha$), for real and shuffled data, of inter-vector distances for the 12 groups sampled from European countries ($\alpha=1,12$; alphabetical order) plus the group sampled across Europe ($\alpha=13$) and the set of uniformly random data ($\alpha=14$).}
\label{tab}
\end{table}
Further important higher-order differences between real and randomized data can be characterized by performing a hierarchical clustering algorithm of the vectors $\{\vec{v}_i\}$, which represents the latter as leaves of a dendrogram where culturally closer individuals have a lower common branching point.
This is shown in fig.\ref{fig_dendro} for real, shuffled, and random data. As one can clearly see, the dendrogram for real data is well structured in sub-branches nested within branches, indicating that cultural space is heterogeneously populated by dense communities of similar individuals, separated by sparsely occupied regions. 
The hierarchical character of this distribution shows that denser regions are iteratively fragmented into denser regions nested within them. 
This peculiar organization indicates that the original distances are (nearly) \emph{ultrametric} \cite{ultrametricity}, i.e. the tree-like representation rendered by the dendrogram is not just an artifact of the clustering algorithm, but a natural property of the data. This means that the height of the first branching point connecting two individuals $i$ and $j$ approximately corresponds to the original distance $d_{ij}$ between $\vec{v}_i$ and $\vec{v}_j$.
By contrast, the dendrograms for shuffled and real data are trivially structured, with no well-defined internal separation between different hierarchical levels. 
In this case the dendrogram is not representative of the original distribution of vectors, and is merely an uninformative outcome of the algorithm which is forcing non-ultrametric data into a tree-like description. In such a situation, the vertical dimension of the dendrogram loses it correspondence with the original inter-vector distances, and provides a highly distorted image of the latter.
Thus, we find that the broader distribution of real distances (with respect to shuffled ones) is implied by the ultrametric structure, characterized on one hand by an increased frequency of both nearby vectors (representing individuals within the same branch of the dendrogram), and on the other hand by an increased frequency of distant ones (representing individuals belonging to different branches). By contrast, shuffled data generate vectors with the same average distance but more uniformly and non-ultrametrically distributed in cultural space.

\section{Local and global levels of influence}
\begin{figure}
\begin{center}
\includegraphics[width=.4\textwidth]{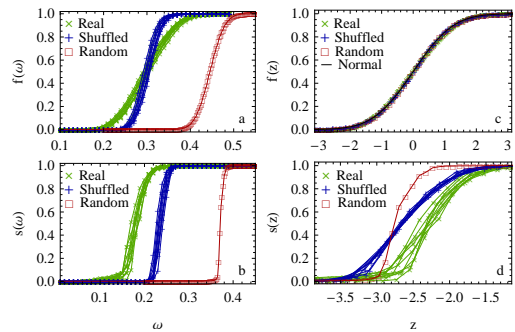}
\end{center}
\caption{Local and global measures of influence in cultural graphs obtained from real and randomized data. \textbf{a)} Fraction of direct influential interactions (link density) $f_\alpha$ as a function of the threshold $\omega$. 
\textbf{b)} Fraction of the largest connected component $s_\alpha$ emerging from indirect influential interactions as a function of $\omega$. 
\textbf{c)} Link density $f_\alpha$ as a function of the rescaled threshold $z\equiv(\omega-\mu_\alpha)/\sigma_\alpha$. The black solid line is the cumulative density function (CDF) of the standard Gaussian distribution. 
\textbf{d)} Fraction of the largest connected component $s_\alpha$ as a function of the rescaled threshold $z$.
\label{fig_fs}}
\end{figure}

The ultrametric hierarchy discussed above has important static and dynamic consequences. As we show in the Appendix, the branches of the dendrogram `cut' horizontally at a distance $\omega$ coincide with the connected components of the $\omega$-dependent cultural graph we defined in the beginning.
It is interesting to study, as a function of $\omega$, the density of links $f_\alpha(\omega)$ (which is nothing but the CDF of the distance distribution) and the size fraction $s_\alpha(\omega)$ of the largest connected component in the cultural graph, which represent a local and a global measure of influence among the individuals of group $\alpha$ respectively. The resulting curves are shown in fig.\ref{fig_fs}a-b. For both quantities, we observe large differences between real and randomized data. 
In particular we find that, for a given value of $\alpha$, real data are characterized by  higher levels of local and global influence than shuffled and random data. 
In order to understand whether the differences among the curves in figs.\ref{fig_fs}a and \ref{fig_fs}b can be simply traced back to overall differences in the average values ($\mu_\alpha$) and variances ($\sigma^2_\alpha$) of the inter-vector distances, in fig.\ref{fig_fs}c-d we show $f_\alpha$ and $s_\alpha$ when plotted as a function of the standardized parameter $z\equiv (\omega-\mu_\alpha)/\sigma_\alpha$. 
We find (see fig.\ref{fig_fs}c) that all the $f_\alpha(z)$ plots collapse onto a single universal curve, which is indistinguishable from the cumulative density function (CDF) of the standard Gaussian distribution $f_0(z)\equiv\int_{-\infty}^z dx(e^{-x^2/2})/\sqrt{2\pi}$.
In other words, in each group $\alpha$ the distances are normally distributed, with group-specific mean $\mu_\alpha$ and variance $\sigma^2_\alpha$. 
This means that all the empirical differences in link density among groups are taken care of after rescaling the distance, so that the control parameter $z$ completely specifies the density of realized cultural channels of any group. 
Thus, if cultural channels were placed uniformly among individuals as in a homogeneous random graph, one would also observe an analogous universal collapse of the $s_\alpha(z)$ curves and of any other topological property. By contrast, as can be seen in fig.\ref{fig_fs}d, this approximately occurs only in the shuffled case, but not for real data. This result indicates again a nontrivial distribution of real cultural vectors, and singles out differences across the sampled groups that are not simply explained in terms of an overall variability. 
In particular, the universality of the link density function observed in fig.\ref{fig_fs}c does not imply a universal structure of connected components, due to correlations between pairs of edges generated by the correlations between distances. 
In other words, even after standardizing the local level of mutual influence, real data continue to differ significantly in their global  level of influence.
Therefore any process which depends on the cultural distance between individuals might have very different global outcomes even when taking place on locally identical structures. 

All the above results show that even randomly sampled individuals (as the ones in our database) are not characterized by uniformly random cultural vectors. \footnote{Importantly, the nature of the survey -- which ensures that the sampled individuals are not neighbours in a social network -- allows us to conclude that the observed correlations are not merely an effect of having selected individuals who have already interacted and thus culturally converged. 
For the same reason, we selected one early year (1992) in the database when extensive online interactions were not possible yet, so that the individuals in our analysis have never interacted, either online or offline.}
While this is not surprising, the peculiar hierarchical distribution implied by empirical inter-opinion correlations is highly nontrivial, and unpredictable \emph{a priori}. 
Since, as we show in what follows, the dynamics of opinions and culture is strongly dependent on initial conditions, it is important to investigate how the predictions of popular models change when the empirically observed data are considered as the starting configuration, rather than the ordinarily postulated \cite{axelrod,De_Sanctis_and_Galla_2007,Flache_and_Macy_2006,castellanoreview} random (or even only uncorrelated) cultural vectors.
To this end, in what follows we study how the empirically observed ultrametric structure affects the predictions of models simulating both short-term and long-term dynamics, with a particular interest in exploring the effects on the coexistence of cultural heterogeneity and collective social phenomena.  
Rather than considering one or more (unavoidably arbitrary) specifications of possible social networks that ideally start connecting the (initially non-interacting) individuals of a group, we choose to establish a unique upper bound for the achievable level of influence, where the social network is virtually replaced by a complete graph. This choice corresponds to selecting the maximum level of influence on the social side of the problem, and letting the value of the cultural threshold $\omega$ uniquely determine whether two individuals can influence each other ($d_{ij}<\omega$) or not ($d_{ij}>\omega$) according to the bounded confidence hypothesis.
As discussed in the Appendix, the new possibilities of interactions that have been recently become available on online platforms, where individuals influence each other bypassing social ties, are modifying the traditional scenario and leading us closer to this extreme `complete graph' setting. 
In any case, rather than the dynamical outcome in absolute terms (which strongly depends on the specification of the underlying network), our main interest is the comparison, on the same network, of the outcome implied by real cultural vectors with that implied by randomized data.

\section{Short-term collective social behavior}
We first study the effects of the empirical structure of real opinions on short-term collective social behavior. 
We consider a simple prototypic model where, on short timescales, cultural vectors do not evolve but nonetheless determine the choices that individuals make under the influence of each other. To this end, we extend the Cont-Bouchaud (CB) model \cite{contbouchaud}, originally proposed to model herding effects in financial markets, to a more general `coordination model' which incorporates a dependence on real cultural vectors $\{\vec{v}_i\}$ (see Appendix). 
For each group in our analysis, we consider a situation where individuals are asked, for instance in democratic elections, public referenda, financial markets, online surveys, etc., to make a binary choice such as yes/no, buy/sell, approve/reject, left/right etc.
We can represent the choice expressed by the $i$-th individual as $\phi_i=\pm 1$.
The effects of mutual influence and bounded confidence are modeled by allowing pairs of individuals whose cultural distance $d_{ij}$ is smaller than a threshold $\omega$ (which is the only parameter of the model) to exchange information before making their choices. 
As a result of this information exchange, we assume that all the agents belonging to the same connected component of the resulting $\omega$-dependent cultural graph (see Appendix) collectively agree on the choice to make. 
If $A$ labels a connected component of the graph, the choice of all agents belonging to $A$ is the same ($\phi_i=\phi_A$ $\forall i\in A$), while different connected components make statistically independent choices. 

\begin{figure}
\begin{center}
\includegraphics[width=.45\textwidth]{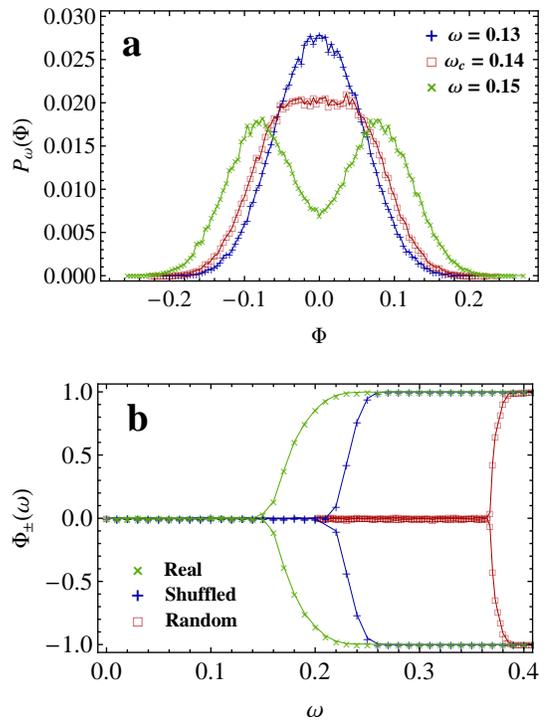}
\end{center}
\caption{At a critical confidence level, a spontaneous breaking of choice symmetry occurs. \textbf{a)} When our `coordination model' is simulated on real data (in the example shown, the Germany group), we observe an abrupt change in the probability $P_\omega(\Phi)$ of a collective consensus at a critical confidence value $\omega_c$. For $\omega<\omega_c$, individual choices are uncorrelated and sum up to a vanishing global outcome $\Phi=0$, at which $P_\omega(\Phi)$ has a single peak. For $\omega>\omega_c$, local interactions result in global correlations that spread across the entire system, and a macroscopically coordinated output, whose probability is peaked about the two nonzero values $\Phi_\pm(\omega)$, emerges. Right at $\omega=\omega_c$, $P_\omega(\Phi)$ displays a flat region typical of critical phenomena. \textbf{b)} The most probable value of $\Phi$ is the order parameter of the phase transition. For $\omega<\omega_c$ it is vanishing, while for $\omega>\omega_c$  it branches into the two symmetric values $\Phi_\pm(\omega)$. In addition to the results for real Germany data, here we also show the results for shuffled and random data. Real data always have a lower critical threshold than randomized data, indicating an enhanced possibility to behave collectively. All the other groups show the same behavior.
\label{fig_critical}}
\end{figure}

The overall outcome of the process (e.g. the result of the survey/referendum/election) is the sum of individual preferences, and can be quantified by the average choice
\begin{equation}
\Phi=\frac{1}{N}\sum_{i=1}^N\phi_i=\frac{1}{N}\sum_A S_A\phi_A
\label{eq_Phi}
\end{equation}
where the second sum runs over all connected components, and $S_A$ is the size of component $A$. The sign of $\Phi$ reflects the choice of the majority, and a key property characterizing the outcome of the model is the probability $P_{\omega}(\Phi)$ that the average choice takes the particular value $\Phi$, for a given value of $\omega$. 
Following the procedure described in Appendix, we computed $P_{\omega}(\Phi)$ for various values of $\omega$ (from $\omega=0$ to $\omega=1$ in increments of $0.01$) and for all the 13 groups in our dataset (both real and shuffled), plus the completely random set.
In fig.\ref{fig_critical}a we report the results for real Germany data. As can be seen, there exists a critical value $\omega_c$ (in the case shown, $\omega_c=0.14\pm 0.01$) such that, for $\omega<\omega_c$, $P_\omega(\Phi)$ is symmetric about zero (as for $\omega=0$) and, for $\omega>\omega_c$, $P_\omega(\Phi)$ has two symmetric peaks. 
Right at $\omega=\omega_c$, $P_\omega(\Phi)$ displays a flattened region. This behavior is typical of symmetry-breaking phase transitions. Here the order parameter of the transition is the most probable value(s) $\Phi_\pm$ of $\Phi$: for $\omega<\omega_c$ one has $\Phi_+(\omega)=\Phi_-(\omega)=0$, while for $\omega>\omega_c$ one has two symmetric values $\Phi_-(\omega)<0<\Phi_+(\omega)$ with $\Phi_-(\omega)=-\Phi_+(\omega)$. This is shown in fig.\ref{fig_critical}b, where we also plot the behavior for shuffled and random data. 
\footnote{Note that, technically, the critical threshold is defined only for infinite systems, and most of the methods available to measure its exact value make use of an ideal thermodynamic limit. However, here we are interested in determining the parameter value that drastically changes the behavior of an intrinsically finite system. An extrapolation to infinite individuals would make any reference to the outcome of real social processes vanish. By contrast, the above method has an easily interpretable meaning even in the finite case. In the infinite size limit, it would yield the exact value of the critical threshold (see Appendix).}

\begin{figure}
\begin{center}
\includegraphics[width=.45\textwidth]{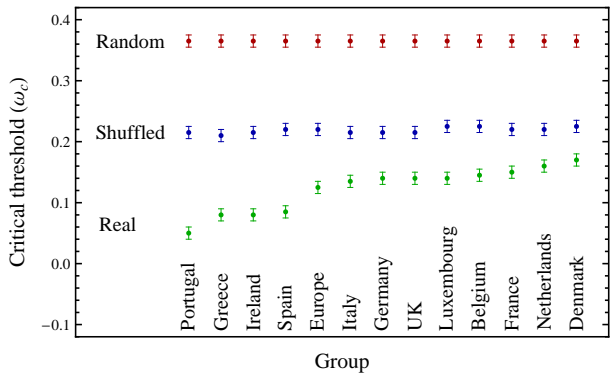}
\end{center}
\caption{Critical thresholds representing resistance to collective behavior of the sampled groups, for real, shuffled and random data. 
For real data, the values of $\omega_c$ are not consistent with each other within the error bars, indicating different resistances to coordination across the sampled groups. 
They are however always smaller than randomized data, indicating that the empirical ultrametric distribution in cultural space systematically facilitates the onset of collective behavior.}
\label{fig_thresholds}
\end{figure}

This analysis allows to measure the critical thresholds $\omega_c$ for all the groups we considered (see fig.\ref{fig_thresholds}). 
Note that smaller (larger) values of $\omega_c$ require smaller (larger) levels of influence between individuals in order to trigger collective behavior.
Therefore $\omega_c$ represents a novel measure of the resistance of a social group to act collectively. 
Importantly, we found that the thresholds for shuffled data are always larger than those for real data (see fig.\ref{fig_thresholds}), and the ones for random data are even larger.
This shows that empirical inter-opinion correlations, which are responsible for the ultrametric distribution of individuals in cultural space, strongly facilitate collective social behavior by systematically lowering the resistance to coordination. 
While for shuffled and random data all thresholds are equal within errors, the entity of the enhancement of collective behavior in real data varies significantly across the sampled groups and determines non-universal values of the  thresholds. 
In general, $\omega_c$ also represents a fundamental threshold for any dynamical process dependent on cultural data. Any mechanism taking place within a cultural distance smaller than $\omega_c$ will not propagate to the whole network, while if the interaction range is larger than $\omega_c$ the information can percolate the entire system. 

This simple model indicates that, depending on the local interaction range, individual differences can either be `averaged out' and disappear at the macroscopic level or give rise to a collectively coordinated behaviour. 
Understanding this transition in real societies is one of the fundamental open questions of modern social science \cite{socialatom}. In economics, this problem is related to whether it is legitimate to use the concept of `representative agent' as an idealized individual that makes the average choice of the society. Contributions from different scientific communities to this ongoing debate gave very different points of view about the subject. 
Our simplified model, when simulated on real data, suggests that both regimes are possible, and that a simple local parameter can trigger very different global outcomes. In particular, the variance of the collective outcome, which grows with the separation between the peaks of $P_\omega(\Phi)$, can either decay to zero or be amplified macroscopically. 
In the latter case, the final outcome is collective (or, in the case of votes, democratic) in a `strong' sense: a large portion of the population makes a coordinated action, and the majority's choice reflects a truly collective consensus. 
By contrast, in the former case the result is collective or democratic in a `weak' sense: there are several groups of people making different choices, and the majority's action does not reflect a collective agreement, but rather a statistical fluctuation about a 50\%-50\% tie \cite{galam}. These considerations indicate that a good measure of the level of collective behavior achievable for a given value of $\omega$ is the width of $P_\omega(\Phi)$. Thus we can define the standard deviation
\begin{equation}
C(\omega)\equiv \sigma_\omega(\Phi)=\sqrt{\sum_A\left(\frac{S_A}{N}\right)_\omega^2}
\label{eq_C}
\end{equation}
as a measure of short-term social \emph{coordination}.
The latter equality in the above formula (see Appendix for a rigorous proof) states that, intriguingly, $C(\omega)$ is uniquely determined by the sizes $\{S_A\}$ of the connected components of the underlying cultural graph obtained for that particular value of $\omega$, and is therefore actually independent of the dynamical model considered. 
This quantity will be useful in what follows. Note that $C(\omega)$ ranges between $0$ and $1$. If $\omega=0$ (no coordination), its value is $C(\omega)\simeq 1/\sqrt{N}$ (as follows from the Central Limit Theorem) and vanishes for large $N$. At the opposite extreme, if $\omega=1$ (complete coordination) then $C(\omega)=1$ which is as large as the standard deviation of the individual choice $\phi_i$, and remains finite when $N\to\infty$.

\section{Long-term cultural diversity} 
We now take an evolutionary perspective and focus on a longer temporal scale over which the cultural vectors  themselves can change. In this case we use a modified version \cite{Flache_and_Macy_2006} of the popular Axelrod model \cite{axelrod}, which is designed to simulate the evolution of vectors of cultural traits on social networks, again in a way that real data can enter into the model.\footnote{Unfortunately, Eurobarometer data are available on a low-frequency basis (generally, every 2 years) and the surveyed individuals (and in some cases the questions themselves) are different from time to time. This does not allow us to measure the actual evolution of the observed opinions over time. The present study suggests the desirability of designing future surveys and experiments that allow to keep track of the evolution of cultural information explicitly. Nonetheless, the dataset offers the unprecedented possibility to simulate the evolution of empirically observed opinions in groups sampled from the real world.}
A detailed discussion of the model can be found in the Appendix.
In an elementary time-step, two individuals $i$ and $j$ belonging to the same sampled group are randomly selected. 
If the generalized overlap $o_{ij}\equiv 1-d_{ij}$ (where $d_{ij}$ is the above-defined distance between the vectors $\vec{v}_i$ and $\vec{v}_j$) is smaller than or equal to $\theta$, no interaction takes place. Otherwise, with probability equal to $o_{ij}$,
the two individuals interact: a component $v_j^{(k)}$, chosen randomly among the components where $\vec{v}_i$ and $\vec{v}_j$ differ, is changed and set equal to $i$'s corresponding component: $v_j^{(k)}=v_i^{(k)}$. Otherwise nothing happens, and two other individuals are selected. 
These rules implement the two basic mechanisms of social influence (interacting actors tend to converge culturally) and homophily (similar individuals interact more frequently).
Note that, in line with our previous analysis, we are assuming that every pair of individuals can interact, i.e. the underlying social network is a complete graph. 
In the allowed final configurations, any two cultural vectors are either completely identical or separated by a distance larger than $\omega\equiv 1-\theta$, and the average $\langle N_D\rangle_\omega$ (over many realizations) of the number $N_D$ of distinct vectors in the final stage, or equivalently the  fraction 
\begin{equation}
D(\omega)\equiv\frac{\langle N_D\rangle_\omega}{N}
\label{eq_D}
\end{equation}
is a convenient way to measure the long-term cultural \emph{diversity} as a function of $\omega$. 

\begin{figure}
\begin{center}
\includegraphics[width=.5\textwidth]{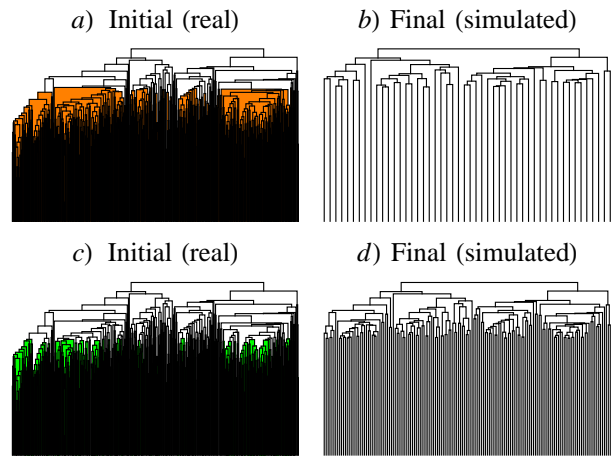}
\end{center}
\caption{The hierarchical structure implied by inter-opinion correlations constrains cultural evolution. 
\textbf{a)} The real, hierarchically organized cultural vectors for the Germany group (the same as shown in fig.\ref{fig_dendro}a) are considered as the initial state of the modified Axelrod model, and a confidence level (corresponding to the horizontal line below which the shaded region originates) is imposed. \textbf{b)} Due to ultrametricity, in the corresponding final state of the model all the individuals within a common shaded branch in the initial dendrogram collapse to the same cultural vector, with negligible effects on the upper part of the dendrogram. \textbf{c)} The same initial state as above is considered, but a lower confidence level is imposed. \textbf{d)} Correspondingly, the final state of the model consists of a larger number of distinct cultural vectors, each containing on average less individuals with collapsed vectors. Thus the number of distinct final vectors (the leaves of the final dendrogram) coincides with the number of branches intersecting the horizontal line in the initial dendrogram. If shuffled or random opinions are taken as the initial state of the model (not shown), this is no longer true since the convergence of cultural vectors also affects the dendrogram's structure above the horizontal line, signalling a lack of ultrametricity.
\label{fig_dendro4}}
\end{figure}

We ran several realizations of the model by taking both real and randomized cultural vectors $\{\vec{v}_i\}$ as the starting configuration.
As we show in the Appendix, we find that real data are those that achieve the largest level of long-term cultural heterogeneity (value of $\langle N_D\rangle$). Indeed, for real data the realized value of $\langle N_D\rangle$ is the largest possible ($\langle N_D\rangle\approx N_C$) indicating that cultural convergence is confined within the initial connected components, each of which eventually becomes a single cultural domain. By contrast, in randomized data there are less final cultural domains than initial connected components, indicating that the latter often `merge' into larger cultural domains.
The reason for the remarkably different behavior of real and randomized data is, once again, the ultrametric character of the former.  As we show in fig.\ref{fig_dendro4}, ultrametricity implies that the branches obtained cutting the real-data dendrogram at some value of $\omega$ will collapse into a single cultural vector. This means that the initial structure of the dendrogram above $\omega$ will be `frozen' and unaffected by cultural evolution. This confines cultural convergence locally within the lower branches.
By contrast, in randomized data the lack of ultrametricity implies that branches are not well separated, so that the local convergence of vectors within a branch reduce the separation of the branch itself from nearby branches. Thus in this case branches are unstable, and often merge modifying the entire structure globally.

\begin{figure}
\begin{center}
\includegraphics[width=.45\textwidth]{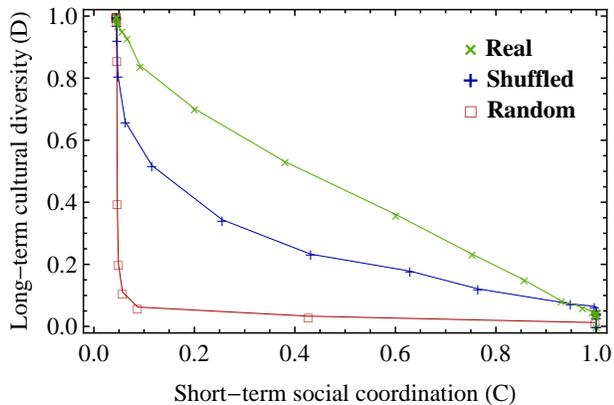}
\end{center}
\caption{Phase diagram summarizing our results. The long-term cultural diversity $D$ is shown as a function of short-term social coordination $C$ for real, shuffled and random data. If random cultural vectors are considered, cultural heterogeneity and  collective behavior are mutually excluding: one has either $D\approx 0$ or $C\approx 0$. This approximately corresponds to the traditional situation explored when considering a random graph of interaction among individuals.
By contrast, real cultural vectors allow high simultaneous levels of short-term coordination and long-term diversity, including the approximately balanced regime $C\approx D\approx 1/2$.
Shuffled data follow an intermediate curve, showing that the heterogeneous frequencies of real opinions and the correlations among the latter both play a significant role in enhancing the coexistence of diversity and coordination in the real world.
\label{fig_diagram}}
\end{figure}

We can combine the above findings with our previous results about collective behavior. In particular, given a group of individuals, we can measure both the short-term social coordination $C(\omega)$ defined in eq.(\ref{eq_C}) and the long-term cultural diversity $D(\omega)$ defined in eq.(\ref{eq_D}) for various values of $\omega$. Then we can plot $D(\omega)$ versus the value $C(\omega)$ obtained for the same $\omega$, as in fig.\ref{fig_diagram}.
If we look at random data, we retrieve the naive result that the coexistence of cultural heterogeneity and social collective behavior is impossible, since we have either $D\approx 0$ or $C\approx 0$.
Note that a cultural graph defined among random vectors is approximately equivalent to a random graph, whose density is completely determined by $\omega$ through the relation shown in fig.\ref{fig_fs}a. Therefore the results we show for random vectors coincide with the standard results that would be obtained by simulating the Cont-Bouchaud and Axelrod models on random graphs, for various density values.
By contrast, we find that real cultural vectors allow high simultaneous levels of short-term coordination and long-term diversity, including the approximately balanced regime $C\approx D\approx 1/2$ which in the case shown is achieved for the moderate influence level $\omega\approx 0.17$.
Shuffled data follow an intermediate curve, showing that the heterogeneous frequencies of real opinions and the correlations among the latter both play a significant role in enhancing the coexistence of diversity and coordination.
Thus, surprisingly, we find that empirical hierarchical correlations simultaneously enhance collective behavior and sustain cultural heterogeneity. While the incompatibility of these two phenomena holds for randomized data, which represent the usual specification of dynamical models, it is violated by real data.
This remarkable result highlights the scarce predictive power of models that consider random specifications, and shows the importance of empirical analyses of high-dimensional cultural vectors, which offer the unprecedented possibility to explore cross-correlations among opinions and their consequences.

\section{Conclusions}
By using a large detailed dataset, we have characterized the empirical properties of the large-scale distribution of individuals in multidimensional cultural space.
We found that real inter-opinion correlations organize individuals hierarchically and ultrametrically in cultural space, a result which is not retrieved when randomized or one-dimensional opinions are considered. 
These properties strongly determine the exploitable network of interactions that is expected to arise as a result of the bounded confidence hypothesis, according to which individuals are only influenced by culturally similar peers. 
Ultrametricity has profound and nontrivial consequences on short- and long-term cultural dynamics. In the short term, if one assumes that consensus can be reached by individuals with sufficiently similar opinions, we found the existence of a symmetry-breaking phase transition where collective behavior arises out of purely local interactions. 
The critical threshold of this transition is remarkably lower in real data than in randomized cases, indicating that ultrametricity enhances short-term collective behavior.
However, in the long term the same ultrametric property suppresses cultural convergence by restricting it withing disjoint domains, implying a strong sensitivity to initial conditions. 
These opposite effects imply that, whereas in random data the coexistence of short-term coordination and long-term diversity is unfeasible, in real data it is strongly enhanced and can be achieved in a broad region of parameter space.
Thus the apparent paradox of the coexistence of short-term collective social behavior and long-term cultural diversity might have, as a simple and parsimonious explanation, the empirically observed hierarchical distribution of individuals in cultural space.

\appendix
\section{Social networks and cultural graphs: the online shift}
Here we discuss how social influence, homophily and bounded confidence are believed to affect social dynamics, and how offline and online environments are expected to contribute differently to the process.
As we mentioned in the main text, the bounded confidence hypothesis states that two individuals are potentially influenced by each other only if the distance between their cultural variables (or vectors) is smaller than a certain threshold $\omega$.
Thus, while social influence takes place on a social network connecting individuals, bounded confidence involves a different graph, that we denote as the \emph{cultural graph}, where pairs of individuals separated by a distance smaller than the confidence $\omega$ are connected by `cultural channels', irrespective of whether they are neighbors in a social network.
When combined together, these hypotheses imply that the actual network of interactions is given by the overlap between social ties and (confidence-dependent) cultural channels. If $s_{ij}$ and $c_{ij}(\omega)$ denote the elements of the adjacency matrix of the social and cultural network respectively (equal to 1 if a link between vertices $i$ and $j$ is there, and 0 otherwise), the overlap network is described by an adjacency matrix with entries $a_{ij}(\omega)= s_{ij}c_{ij}(\omega)$, where $c_{ij}(\omega)=1$ if $d_{ij}<\omega$ and $c_{ij}(\omega)=0$ if $d_{ij}>\omega$. 

Models that simulate the evolution of societies must therefore be complemented by empirical analyses characterizing both social networks and cultural graphs. However, while a huge literature is devoted to the study of real networks formed by social ties (using both traditional small-scale surveys \cite{wasserman} and more recently large-scale communication data \cite{phonecalls,email}), little is known about the empirical properties of those formed by cultural channels. As a consequence, when considering models of opinion dynamics, social networks have been so far considered as proxies for the actual interaction graphs, i.e. $a_{ij}\approx s_{ij}$, which amounts to assume $c_{ij}\approx 1$ (or equivalently an unbounded confidence $\omega=+\infty$) for all pairs of individuals. This assumption is helpful in the traditional offline situation where social ties are expected to be dominant over cultural channels, for instance when social groups are formed independently of the cultural traits of people (e.g. acquaintances made in public schools). However, it prevents our understanding of the opposite extreme, i.e. when people look for culturally similar peers, that they do not know initially, to interact with (for instance by joining an open discussion group on a focused topic of interest). In such a situation the empirical knowledge of $c_{ij}(\omega)$ may become even more important than that of $s_{ij}$. 
While these non-social interactions used to represent only a secondary means of opinion diffusion up to some years ago, they are now becoming more and more pervasive as novel electronic platforms are being developed and used at a large scale \cite{felix_online,onlinepopularity,footprints,computationalsocialscience}. Indeed, in recent years many new possibilities of interaction have rapidly emerged, such as the exchange of opinions through the WWW (on-line forums, blogs, discussion groups, etc.) and other media. In these platforms, people already interested in a topic search new peers (who they 
may never meet physically afterwards) to discuss with, experiencing novel ways by which opinions can interact. These interactions can be even stronger than those occurring through direct knowledge, as virtual communities gather together people with oriented and focused interests, who often recognize each other as the most natural and qualified peers to share ideas with, even if no direct knowledge exists between them. 

The above picture suggests an extension of the idea of homophily, according to which `likes attract', to an online setting. 
When these additional possibilities of interaction are considered, cultural channels may become the dominant means of interaction, as people with similar interests are more likely to access the same platform(s) and exchange opinions more frequently than culturally dissimilar individuals. In this opposite extreme, the `social' network through which opinions can in principle interact is virtually replaced by a complete graph (i.e., $s_{ij}=1$ $\forall i,j$) where everybody is connected to everyone else (in social space, the interaction becomes infinite-ranged). Cultural graphs therefore become a natural proxy for the actual interaction graph: $a_{ij}(\omega)\approx c_{ij}(\omega)$. 
Despite its increasing importance, our  understanding of this novel type of long-range opinion dynamics, which bypasses social ties and is dominated by the structure of real opinions and cultural graphs, is incomplete.
Our analysis bridges this gap by using real data about the opinions, beliefs and attitudes of thousands of Europeans to produce cultural graphs where we can investigate the outcome of infinite-range, online-like dynamics on both short and long time scales.

\section{Definition of cultural vectors and distances}
Here we describe how we reconstructed real multidimensional opinions from empirical data.
We note that, for our analysis to be insightful, we need to access the opinions of a set of real individuals who do not know each other and are well separated socially. This is essential in order to separate the purely cultural dynamical effects from ordinary social influence effects (producing convergent cultural traits) that might be already present in the data if the individuals are sampled nearby in a social network. 
For this reason, we focused on a large dataset that is specifically designed to survey a number of beliefs and opinions across Europe, and based on standard sampling protocols ensuring that individuals are selected avoiding the bias due to (among other factors) social closeness. This dataset \cite{dataset,eurobarometer}, an official release of the Eurobarometer project \cite{agnes1,agnes2}, reports (in its 1992 snapshot\footnote{The reason for choosing this early year in the dataset is to rule out the possibility, besides offline social interactions, of online interactions among the sampled individuals.}) the results of face-to-face interviews where about 13,000 individuals across 12 European countries\footnote{In alphabetical order: Belgium, Denmark, France, Germany, Greece, Ireland, Italy, Luxembourg, the Netherlands, Portugal, Spain, UK.}) were asked to fill a questionnaire containing several multiple-choice questions. These questions were designed to capture a range of individual beliefs and attitudes towards various scientific topics, thus surveying the `Public Understanding of Science' across Europe. 
Besides probing the general level of scientific awareness in the individuals, the questionnaire focused on issues that were considered `hot topics' in relation to European integration, e.g. the introduction of novel biotechnologies, the role of scientists in the dissemination of their research results, various bioethical questions, etc. 
The database is invaluable in order to study the real multidimensional organization of opinions, as well as its dynamical consequences. 
In particular, it allows us to establish a previously unavailable empirical reference for theoretical models, such as the aforementioned one proposed by Axelrod, where individuals are represented as vectors of cultural traits evolving through the interaction with peers.

Raw data, originally arranged in a SPSS spreadsheet file of $N=13,000$ rows where row $i$ reported the (numerically coded) answers of the $i$-th individual to $F=161$ multiple-choice questions, were transformed into $N$  $F$-dimensional vectors $\{\vec{v}_i\}$. 
Each component $v_i^{(k)}$ of these vectors was given a value such that the corresponding contribution $d^{(k)}_{ij}$ to the overall distance
\begin{equation}
d_{ij}=\frac{1}{F}\sum_{k=1}^F d_{ij}^{(k)}
\label{eq_distance}
\end{equation}
is in the range $[0,1]$, with $d^{(k)}_{ij}=0$ representing $i$ and $j$ giving an identical answer to question $k$ and $d^{(k)}_{ij}=1$ representing $i$ and $j$ giving opposite answers. 
For `metric' questions, where answers were possible in an equally spaced scale of $Q_k$ possibilities, the maximum information is retained by mapping the original answers to the possible values $$v_i^{(k)}=0,\frac{1}{Q_k-1},\frac{2}{Q_k-1},\dots,1$$ and defining $d^{(k)}_{ij}\equiv |v_i^{(k)}-v_j^{(k)}|$. 
For non-metric questions (associated to $Q_k$ unordered possible alternatives), 
`opposite answers' simply means `different answers', and we therefore mapped the possible values of $v_i^{(k)}$  to $Q_k$ arbitrary symbols and defined $d^{(k)}_{ij}\equiv 1$  if $v_i^{(k)}=v_j^{(k)}$   and  $d^{(k)}_{ij}\equiv 0$ otherwise. If all questions were non-metric, this choice would be equivalent to $d_{ij}=1-o_{ij}/F$ where  $o_{ij}$ is the overlap (number of components with identical value) between $\vec{v}_i$  and  $\vec{v}_j$, a commonly used notion of cultural similarity \cite{axelrod} as mentioned in the main text. Note that for binary answers (such as `yes'/`no') the metric and non-metric definitions coincide. 

Each individual belongs to one of the 12 European countries in the dataset. This information allowed us to generate groups of individuals sampled either from the same country, or from different ones. 
We labeled different groups of individuals with different Greek letters ($\alpha,\beta,\dots$), and use the notation $i\in \alpha$ to indicate an individual $i$ belonging to group $\alpha$.
In order to deal with samples of equal size, we selected $N=500$ individuals per country\footnote{The dataset contains a sample of approximately 1,000 individuals per country, except Luxembourg (500 individuals) and Germany (2000 individuals aggregating two sets of 1000 individuals, one for East Germany and one for West Germany according to the division existing in 1992). Thus 500 is the maximum size we can take if we want samples of equal size without discarding any country.} and generated 12 groups accordingly. This reproduces a situation where, for instance, people with the same language join a medium-sized online discussion group mediating electronic, non-social interactions mentioned above. Similarly, it mimics individuals that are physically `put together' to participate to a discussion group or to a social experiment. Finally, this choice allows to establish an upper bound for the cultural homogeneity predicted by the `traditional' dynamics taking place on any possible social network connecting the same individuals. 
We also generated an additional thirteenth group with $N=500$ individuals sampled from all the 12 European countries, and denoted it as the \emph{Europe} group. This reproduces a situation analogous to the one described above, but where several individuals across Europe can form a group irrespective of their nationalities, e.g. using a common language such as English.

\section{Measuring inter-opinion correlations}
Our multidimensional data allowed us to investigate cross-correlations among opinions, an information which is not available in one-dimensional studies. 
We measured the covariance matrix between $d^{(k)}_{ij}/F$ and $d^{(l)}_{ij}/F$, whose entries read 
\begin{equation}
\sigma^{(k,l)}_\alpha\equiv\frac{
\langle d^{(k)}_{ij}d^{(l)}_{ij}\rangle_{ij\in\alpha}-
\langle d^{(k)}_{ij}\rangle_{ij\in\alpha}
\langle d^{(l)}_{ij}\rangle_{ij\in\alpha}}{F^2}
\end{equation}
where $\langle\cdot\rangle_{ij\in\alpha}$ denotes an average over all pairs of individuals in group $\alpha$. 
From the above matrix it is possible to obtain the inter-opinion correlation matrix, whose entries read
\begin{equation}
\rho^{(k,l)}_\alpha\equiv
\frac{\sigma^{(k,l)}_\alpha}{\sigma^{(k,k)}_\alpha \sigma^{(l,l)}_\alpha}
\end{equation}
and range between $-1$ (perfect anticorrelation) and $1$ (perfect correlation).
\begin{figure}
\begin{center}
\includegraphics[width=.49\textwidth]{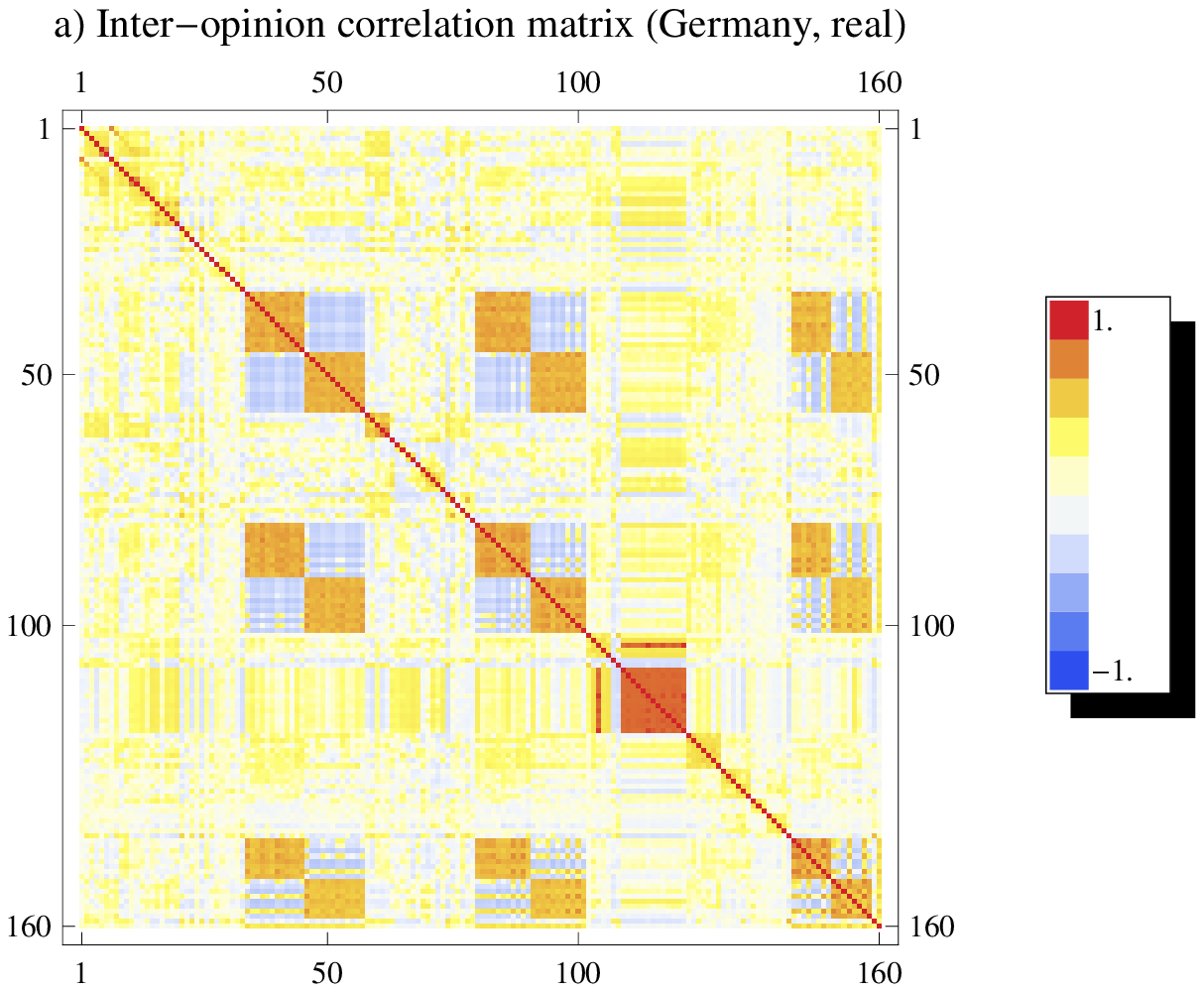}
\includegraphics[width=.49\textwidth]{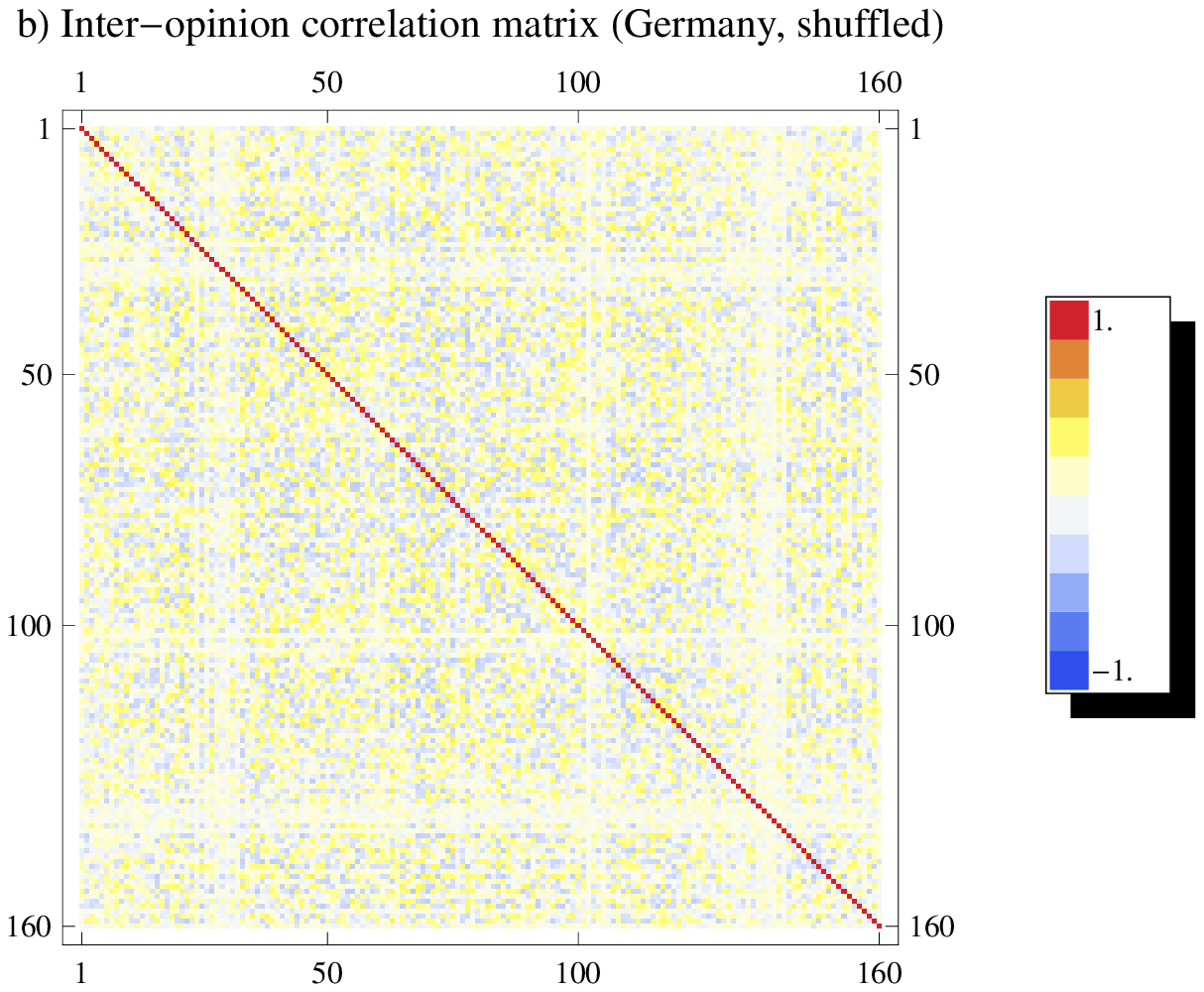}
\end{center}
\caption{
Colour plots of Inter-opinion correlation matrices $\rho^{(k,l)}_\alpha$ for the Germany group. \textbf{a)} real data. \textbf{b)} shuffled data. 
\label{fig_matrices}}
\end{figure}
A colour plot of the correlation matrix is shown in fig.\ref{fig_matrices}a for the group sampled from Germany data, which will remain a reference case throughout our analysis (similar results are found for all the other sampled groups).
As can be seen, there are several pairs of opinions $(k,l)$ characterized by strong positive correlations ($\rho^{(k,l)}_\alpha\approx 1$), but only a few pairs with weak negative correlations ($\rho^{(k,l)}_\alpha\lesssim 0$). 
The predominance of positive correlations indicates that, in general, two individuals $i$ and $j$ giving similar/different answers to question $k$ (small/large $d_{ij}^{(k)}$) tend to give similar/different answers to question $l$ as well (small/large $d_{ij}^{(l)}$), while the opposite outcome (small $d_{ij}^{(k)}$ and large $d_{ij}^{(l)}$) occurs much less frequently.
As we show in fig.\ref{fig_matrices}b, the inter-opinion correlation matrix for shuffled data lacks any structure, indicating the absence of statistically significant correlations (obviously, the same is true for random data, not shown).

The elements of the covariance matrix determine the intra-group variance
\begin{equation}
\sigma^2_\alpha\equiv
\langle d_{ij}^2\rangle_{ij\in\alpha}-\langle d_{ij}\rangle^2_{ij\in\alpha}=
\sum_{k,l}\sigma^{(k,l)}_\alpha
\end{equation}
Note that for shuffled data we have
\begin{equation}
\sigma^2_{\alpha,shuffled}
=\sum_{k}\sigma^{(k,k)}_{\alpha,shuffled}
=\sum_{k}\sigma^{(k,k)}_{\alpha,real} 
\end{equation}
since $\sigma^{(k,k)}_{\alpha,shuffled}=\sigma^{(k,k)}_{\alpha,real}$ and $\sigma^{(k,l)}_{\alpha,shuffled}=0$ for $k\ne l$, while for real data we have
\begin{equation}
\sigma^2_{\alpha,real}=
\sigma^2_{\alpha,shuffled}+\sum_{k\ne l}\sigma^{(k,l)}_{\alpha,real}>\sigma^2_{\alpha,shuffled}
\end{equation}
where the last inequality comes from the observed positivity of $\sum_{k\ne l}\sigma^{(k,l)}_{\alpha,real}$.
Therefore, even if the distribution of inter-vector distances has the same average value in real and shuffled data, real distances are more broadly distributed than shuffled ones. 


\section{Link density and largest connected components}
The ultrametricity of real inter-vector distances implies that if we `cut' the dendrogram of cultural vectors at some height $\omega$ we obtain a set of disconnected branches, within which individuals are separated by a cultural distance smaller than $\omega$, and across which individuals are separated by a distance larger than $\omega$. 
In other words, we obtain the connected components of the cultural graph defined by linking pairs of individuals separated by a distance lower than a certain threshold  $\omega$.
The concept of bounded confidence implies that individuals belonging to different connected components of the cultural graph, even if linked by a social tie, cannot interact. 
Therefore in a fragmented cultural graph information (intended as mutual influence) can only diffuse locally.
A necessary condition in order to have a global spread of information is that cultural channels form a giant connected component spanning (a finite fraction of) the $N$ individuals in a given group. 
The fraction $s_\alpha(\omega)$ of vertices spanned by the largest connected component for a given value of $\omega$ represents an upper bound for the fraction of individuals in group $\alpha$ that can mutually influence each other, through either direct or indirect interactions, if the confidence threshold is set to $\omega$.
Therefore $s_\alpha(\omega)$ is a global measure of potential influence, capturing how local interactions combine together at a large-scale level. 
It is also important to consider a purely local measure of influence, i.e. the average probability that any two individuals can influence each other through a direct interaction. To this end, we also measure the density $f_\alpha(\omega)$ of realized cultural channels, i.e. the fraction of pairs of individuals in group $\alpha$ closer than $\omega$. Note that $f_\alpha(\omega)$ is simply the cumulative density function (CDF) of the distance distribution, i.e. it counts how many pairs of individuals are at distance smaller than $\omega$. For ultrametric data, it also coincides with the fraction of pairs of cultural vectors within the same connected branches, when the dendrogram is cut at the height $\omega$ as discussed above. 
By contrast, for randomized data both $f_\alpha(\omega)$ and $s_\alpha(\omega)$ are no longer in relation with the structure of the dendrogram `cut' at a given point in the vertical dimension, since the latter does not represent the original distances, due to the lack of ultrametricity.

\section{The modified Cont-Bouchaud model}
In its simplest formulation, the Cont-Bouchaud (CB) model \cite{contbouchaud} considers a population of individuals (in the financial jargon, agents) that can make a binary choice between buying or selling an asset traded in the market. 
We can represent the choice expressed by the $i$-th agent as $\phi_i=\pm 1$.
Binary choices are the simplest possibility considered also in other  models of social processes, such as voting dynamics \cite{castellanoreview}. 
In the CB model, the effects of mutual influence are modeled by introducing a random graph through which agents can exchange information before making their choices. 
As a result of this information exchange, all the agents belonging to the same connected component are assumed to collectively agree on the choice to make. 
Therefore, if $A$ labels a connected component of the graph, the choice of all agents belonging to $A$ is the same ($\phi_i=\phi_A$ $\forall i\in A$), while different connected components make statistically independent choices. 
The key result is that the probability distribution of the aggregate choice of all individuals (which in the CB model is the aggregate demand determining the price change of the asset) crucially depends on the topology of the interaction graph.
In particular, if the connection probability $p$ is set at the critical value $p_c\sim N^{-1}$ (for $N\to\infty$) of the phase transition giving rise to the giant connected component, the distribution of the sizes of connected components acquires a power-law form, which in turn implies a power-law distribution of price returns similar to the empirically observed ones.
Other values of $p$ yield different outcomes. In the limit $p=0$ (empty graph) all agents make independent choices and the distribution becomes Gaussian. By contrast, for $p=1$ (complete graph) all agents always make the same choice and the distribution is double-peaked.

In order to study the effects of the nontrivial distribution of individuals in cultural space, we extended the CB model to a more general `coordination model' which incorporates a dependence on real data. 
In particular, we introduce a more realistic mechanism allowing culturally similar agents to express similar preferences. To take this aspect into account 
we assume that each agent is described by a cultural vector $\{\vec{v}_i\}$ and that agents interact, rather than on a random graph defined by a value of $p$, on the cultural graph defined by a value of the confidence $\omega$. 
Again, agents within the same connected component are assumed to make collectively the same choice, while agents belonging to different components express statistically independent preferences.

The overall outcome of the process (e.g. the result of the survey/referendum/election) is the sum of individual preferences, and can be quantified by the average choice
\begin{equation}
\Phi=\frac{1}{N}\sum_{i=1}^N\phi_i
\label{eq_Phiapp}
\end{equation}
whose sign reflects the choice of the majority. 
If the choices of all agents are independent ($\omega=0$), $\Phi$ is the sum of $N$ uncorrelated random variables with finite variance and, as follows from the Central Limit Theorem, normally distributed. In such a case, if the individual binary probabilities are equal (i.e. the events $\phi_i=+1$ and $\phi_i=- 1$ are equiprobable), then the probability $P_\omega(\Phi)$ that the average choice takes the particular value $\Phi$ is symmetric about the most probable value $\Phi=0$. 
If $\omega>0$, $\Phi$ can be rewritten as the sum over different connected components: 
\begin{equation}
\Phi=\frac{1}{N}\sum_A S_A\phi_A
\label{eq_PhiS}
\end{equation}
where now $A$ labels the components, $\phi_A=\pm 1$ is the choice of all actors in component $A$, and $S_A$ is the size of $A$. 
Now, a crucial point is that even if each connected component makes one of the two choices $\phi_A=\pm 1$ with equal probability, and hence the distribution $P_\omega(\Phi)$ is still symmetric about $\Phi=0$, the symmetry breaks spontaneously at the critical threshold $\omega_c$. 
To see this, one can compute $P_\omega(\Phi)$ in the following manner. For each group of individuals in our data, and for a given value of $\omega$ (from $\omega=0$ to $\omega=1$ in increments of $0.01$), we identify the connected components, assign one of the two choices randomly to each of them, and compute the resulting value of $\Phi$. 
We repeat this procedure 500,000 times on each sampled group and measure $P_\omega(\Phi)$ as the normalized histogram of the values obtained. 
The above analysis provides a method to determine with small indeterminacy ($\Delta\omega=0.01$) the threshold value $\omega_c$ for each particular, finite group under study. 
We repeated our analysis on each of the 13 sampled groups (real and randomized) and obtained the corresponding critical thresholds $\omega_c\pm\Delta\omega$.

We found that, while the critical thresholds obtained for shuffled data (and, trivially, also random ones) are consistent with each other, real data feature different critical values. 
This implies that even two randomly sampled social groups (for instance one in Italy and one in Portugal) with the same size and under the same level of mutual influence may evolve to opposite collective states (coordination or heterogeneity) if the critical thresholds for the two groups differ.

We now prove rigorously the equality
\begin{equation}
C(\omega)\equiv \sigma_\omega(\Phi)=\sqrt{\sum_A\left(\frac{S_A}{N}\right)_\omega^2}
\label{eq_Capp}
\end{equation}
which establishes a tight relation between network topology and the level of collective social behavior in the model.
If, for a given value of $\omega$, we denote the expected value of $\Phi$ as $\langle \Phi\rangle_\omega=\sum_\Phi P_\omega(\Phi)\Phi$ and its second moment as $\langle \Phi^2\rangle_\omega=\sum_\Phi P_\omega(\Phi)\Phi^2$, the variance $\sigma_\omega^2(\Phi)$ is defined as
\begin{equation}
\sigma_\omega^2(\Phi)\equiv\langle \Phi^2\rangle_\omega-\langle \Phi\rangle_\omega^2
\end{equation}
For simplicity, in what follows we drop the dependence of all quantities on $\omega$. 
For a fixed value of $\omega$, the sizes of the connected components of the network are given by $\{S_A\}$, and determine the aggregate choice $\Phi$ through eq.(\ref{eq_PhiS}).
Since $\Phi$ is a sum of the random variables $\{S_A\phi_A/N\}$, its variance $\sigma^2(\Phi)$ can be easily expressed as
\begin{equation}
\sigma^2(\Phi)=\sum_{A}\sum_{B}\frac{S_A S_B}{N^2}\sigma_{AB}=\sum_A \frac{S_A^2 \sigma_A^2}{N^2}+\sum_{A}\sum_{B\ne A}\frac{S_A S_B}{N^2}\sigma_{AB}
\end{equation}
where $\sigma_{AB}$ denotes the covariance between the choices $\phi_A$ and $\phi_B$ of two different connected components $A$ and $B$, and $\sigma^2_A$ is the variance of $\phi_A$. Now, since 
\begin{equation} \sigma^2_A\equiv\langle\phi^2_A\rangle-\langle\phi_A\rangle^2=1
\end{equation}
and since different connected components make statistically independent choices, it follows that
\begin{equation}
\sigma_{AB}\equiv \langle \phi_A\phi_B\rangle-\langle \phi_A\rangle\langle\phi_B\rangle =\delta_{AB}\sigma^2_A=\delta_{AB}
\end{equation}
where $\delta_{AB}=1$ if $A=B$ and $\delta_{AB}=0$ if $A\ne B$. Thus the variance of $\Phi$ is simply
\begin{equation}
\sigma^2(\Phi)=\sum_A\frac{S_A^2}{N^2}
\label{eq_proof}
\end{equation}
which proves the last equality in eq.(\ref{eq_Capp}).

Note that for $\omega=0$ there are $N$ connected components of size $S_A=1$ (all vertices are isolated) and therefore $\sigma_0^2(\Phi)=1/N$ (the results of the Central Limit Theorem are recovered). In the opposite limit $\omega=1$, the network is a single connected component of size $S_A=N$, which yields $\sigma_1^2(\Phi)=1$. 
Thus the social coordination $C(\omega)\equiv\sigma_\omega(\Phi)$ varies from $C(0)=1/\sqrt{N}$ (no collective behavior) to $C(1)=1$ (perfect collective behavior). For generic values of $\omega$ note that, since $\sum_A S_A=N$, the expression for $\sigma^2(\Phi)$ in eq.(\ref{eq_proof}) has the form of an inverse participation ratio. This means that $\sigma^2(\Phi)\simeq 1/n$ if the sum over $A$ is dominated by $n$ terms of approximately equal size. In particular, $\sigma^2(\Phi)\simeq 1$ if there is one dominant connected component, while $\sigma^2(\Phi)\simeq 1/N$ if each connected component trivially contains only one vertex. 
This result rephrases the connection between the shape of $P(\Phi)$ and the underlying network topology: when there is no giant component, the width of $P(\Phi)$ is $\sigma(\Phi)\simeq 1/\sqrt{N}\to 0$, while when the giant component is there the width of $P(\Phi)$ has the finite value $\sigma(\Phi)\simeq 1$. 
Thus when there is no giant component there must be a single peak, while the presence of two peaks at finite distance necessarily implies the presence of the giant component. In such a case, $\sigma(\Phi)$  gives an estimate of the separation between the peaks. 

Importantly, these results are valid as $N$ goes to infinity, therefore our method to compute $\omega_c$ as the value marking a spontaneous symmetry breaking (from single-peaked to double-peaked) in the probability $P_\omega(\Phi)$ provides a consistent way to define a `critical' value, which technically is defined only for infinite systems, even for our inherently finite data. For infinite systems, our method would yield the correct value of the critical threshold.
Other finite-size techniques would require assumptions about how topological quantities scale with network size. While for theoretical models (such as the Erd\H{o}s-R\'enyi random graph \cite{guidosbook}) it is possible to derive these assumptions, for real systems this is not possible.

\section{The modified Axelrod model}
In the original version of the Axelrod model, $N$ individuals (in social science jargon, `actors') sitting at the vertices of a social network are represented as vectors of cultural traits (or features), that evolve through discrete steps. In an elementary time-step, an individual $i$ and one of his neighbors (say $j$) are selected. Then the normalized overlap $o_{ij}\in [0,1]$ between their cultural vectors is computed as the fraction of identical components (note that the overlap is related to the cultural distance $d_{ij}$ through $o_{ij}=1-d_{ij}$). With probability equal to $o_{ij}$, the two actors interact: one of $j$'s traits, chosen randomly among the set of traits where $i$ and $j$ differ, is changed and set equal to the corresponding trait of $i$. Otherwise nothing happens, and two other actors are selected. These rules implement the two basic mechanisms of social influence (neighboring actors tend to converge culturally) and homophily (similar individuals interact more frequently). 
The Axelrod model leads to the important conclusion that these two mechanisms do not necessarily reinforce each other leading to a culturally homogeneous society. In fact, the model predicts that diversity is preserved: when two individuals become completely different (zero overlap), they no longer interact. Thus in the allowed final configurations two neighboring actors are either completely identical or completely different, and the society is split into \emph{cultural domains} of identical vectors, with no overlap between adjacent domains. The average $\langle N_D\rangle$ (over many realizations) of the number $N_D$ of different domains in the final stage, or equivalently the  fraction $\langle N_D\rangle/N$, is a convenient way to measure the predicted cultural diversity as a function of the model parameters. 
\begin{figure}
\begin{center}
\includegraphics[width=.48\textwidth]{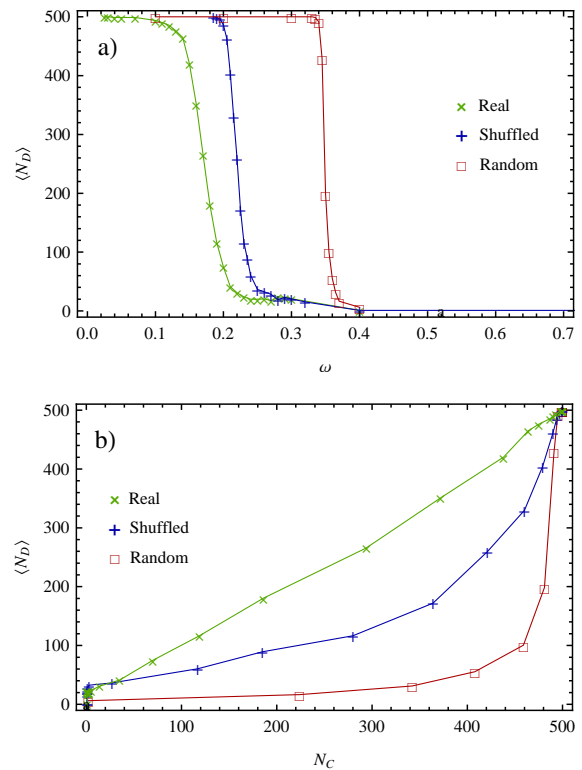}
\end{center}
\caption{The ultrametric properties of real opinions constrain the evolution of cultural convergence even for infinite-range interactions in social space. 
\textbf{a)} The average number $\langle N_D\rangle_\omega$ of cultural domains obtained in the final state of the Axelrod model as a function of $\omega$, when the initial state is given by real, shuffled, and random opinions. 
\textbf{b)} The average number $\langle N_D\rangle$ of final cultural domains versus the number $N_C$ of initial connected components in the cultural graphs, for real, shuffled and random opinions.
\label{fig_axelrod}}
\end{figure}

As for the CB model, for our purposes it is important to incorporate real data into the  Axelrod model. In this case, convenient generalizations have already been proposed. While in the original model traits were non-metric, a modification bringing us closer to real data is the introduction of metric features and the consequent redefinition of cultural distance $d_{ij}$ as a metric distance between cultural vectors \cite{De_Sanctis_and_Galla_2007}. Another important variant introduces the effect of bounded confidence: Flache and Macy \cite{Flache_and_Macy_2006} introduced a threshold $\theta$ such that, if the overlap is smaller than or equal to $\theta$, no interaction takes place. Otherwise, it takes place with probability $o_{ij}$ (the original version of the model is recovered if $\theta=0$). Clearly, $\theta$ has exactly the same meaning as $1-\omega$, where $\omega$ is the confidence we introduced above. The threshold compensates the effect that, as the number of features grows, the presence of completely different pairs of agents becomes unlikely. It also allows to reproduce more realistically the fact that individuals are uninfluenced by each other if they are different enough, not necessarily on each and every opinion they have. \footnote{We also note that it can compensate for the arbitrariness of the number of features appearing in the cultural vectors: two actors with exactly the same vectors may result slightly different if additional features were considered, making the concept of zero overlap not well defined. This is particularly important when dealing with real-world opinions such as our questionnaire data: adding just one more question to the questionnaire cannot change the nature of cultural evolution.}

The above modifications allowed us to use the empirical cultural vectors (rather than commonly assumed uniformly random vectors) as the starting configuration, and study the final diversity predicted by the model. 
In order to simulate long-range online-like dynamics we assumed that the `social' network is a complete graph. 
Since we never found more than one individual with exactly the same cultural vector, the initial number of cultural domains was always $N=500$ for each sampled group. 
In fig.\ref{fig_axelrod}a we report the average number $\langle N_D\rangle_\omega$ of different cultural domains in the final state of the dynamics, as a function of $\omega=1-\theta$ for real, shuffled and random opinions. 
As can be seen, for large values of $\theta$ (small $\omega$) the final fraction of culturally homogeneous domains is finite (at the extreme $\theta=1$ there is no evolution from the starting configuration and the initial vectors remain all distinct), while for small values of $\theta$ (large $\omega$) the same fraction is of order $1/N$ (and vanishes at the extreme $\theta=0$ corresponding to the ordinary Axelrod model). 
This means that the final cultural diversity decreases as $\omega$ increases. With respect to real data, the curved for shuffled and random data are moved rightwards.
Naively, if combined with our previous results, this finding appears to confirm the expectation that larger cultural diversity is only reached in a regime (small $\omega$) where short-term collective behavior is weak or absent, and conversely strong collective behavior can only exist for large values of $\omega$ which suppress cultural heterogeneity in the long run. Moreover, this appears to apply equally to real, shuffled and random data.

However, this conclusion is incorrect. 
In fig.\ref{fig_axelrod}b we show the average number $\langle N_D\rangle$ of final different cultural domains versus the number $N_C$ of initial connected components in the underlying cultural network, both obtained for various values of the threshold $\omega$ and for the three usual cases of real, shuffled and random data.
We find that, for a given value of $N_C$, real data are those that achieve the largest level of long-term cultural heterogeneity (value of $\langle N_D\rangle$). Indeed, for real data the realized value of $\langle N_D\rangle$ is the largest possible\footnote{As we mentioned, in real data ultrametricity implies that the initial connected components are entire branches of the dendrogram, and are therefore completely connected cliques. Since in a clique everyone can interact with everyone else, all individuals in the same component will eventually converge to the same cultural vector. Also note that it is extremely unlikely, for high-dimensional vectors such as the ones in our analysis ($F=161$), that two distinct connected components will end up with the same cultural vector by chance. Thus, for real data, the maximum value of $\langle N_D\rangle$ is $N_C$.} ($\langle N_D\rangle\approx N_C$) indicating that cultural convergence is confined within the initial connected components, each of which eventually becomes a single cultural domain. By contrast, in randomized data there are less final cultural domains than initial connected components, indicating that the latter often `merge' into larger cultural domains.


\end{document}